\documentclass[12pt]{article}

\usepackage{amsmath,amssymb}
\usepackage{graphicx}

\makeatletter

\renewcommand\section{{\setcounter{equation}{0}}
			       \@startsection{section}{1}{\z@}
                                   {-3.5ex \@plus -1ex \@minus -.2ex}
                                   {2.3ex \@plus .2ex}
                                   {\normalfont\large\bfseries}}
\renewcommand\subsection{\@startsection{subsection}{2}{\z@}
                                   {-3.25ex\@plus -1ex \@minus -.2ex}
                                   {1.5ex \@plus .2ex}
                                   {\normalfont\normalsize\bfseries}}
\renewcommand\subsubsection{\@startsection{subsubsection}{3}{\z@}
                                   {-3.25ex\@plus -1ex \@minus -.2ex}
                                   {1.5ex \@plus .2ex}
                                   {\normalfont\normalsize\bfseries}}
\renewcommand\paragraph{\@startsection{paragraph}{4}{\z@}
                                   {3.25ex \@plus1ex \@minus.2ex}
                                   {-1em}
                                   {\normalfont\normalsize\bfseries}}

\makeatother

\newcommand{\beq}{\begin{equation}}
\newcommand{\eeq}{\end{equation}}
\newcommand{\bea}{\begin{eqnarray}}
\newcommand{\eea}{\end{eqnarray}}

\newcommand{\SL}{{\rm SL}}
\newcommand{\SU}{{\rm SU}}

\newcommand{\Sp}{{\rm Sp}}
\newcommand{\Spin}{\rm Spin}
\newcommand{\su}{{\rm su}}

\newcommand{\R}{\mathbb R}
\newcommand{\Z}{\mathbb Z}

\newcommand{\id}{\hbox{1\kern-.27em l}}

\newcommand{\cA}{{\cal A}}

\newcommand{\cC}{{\cal C}}
\newcommand{\cD}{{\cal D}}

\newcommand{\cH}{{\cal H}}

\newcommand{\cN}{{\cal N}}
\newcommand{\cO}{{\cal O}}

\newcommand{\cR}{{\cal R}}

\newcommand{\cT}{{\cal T}}

\begin{document}

\pagestyle{empty}

\begin{center}

\vspace*{30mm}
{\Large  BPS partition functions in $N = 4$ Yang-Mills theory on $T^4$}

\vspace*{30mm}
{\large M{\aa}ns Henningson and Fredrik Ohlsson}

\vspace*{5mm}
Department of Fundamental Physics\\
Chalmers University of Technology\\
S-412 96 G\"oteborg, Sweden\\[3mm]
{\tt mans, fredrik.ohlsson@chalmers.se}     
     
\vspace*{30mm}{\bf Abstract:} 
\end{center}
We consider $N = 4$ Yang-Mills theory on a flat four-torus with the $R$-symmetry current coupled to a flat background connection. The partition function depends on the coupling constant of the theory, but when it is expanded in a power series in the $R$-symmetry connection around the loci at which one of the supersymmetries is unbroken, the constant and linear terms are in fact independent of the coupling constant and can be computed at weak coupling for all non-trivial 't~Hooft fluxes. The case of a trivial 't~Hooft flux is difficult because of infrared problems, but the corresponding terms in the partition function are uniquely determined by $S$-duality.
 
\newpage \pagestyle{plain}

\section{Introduction}
The $N = 4$ maximally supersymmetric Yang-Mills theory in four dimensions possesses a plethora of remarkable properties. The $S$-duality property, which generalizes the strong-weak duality of Montonen and Olive~\cite{Montonen:1977,Goddard:1977}, is among the most studied. Although it remains a conjecture, it is very well established at this point, through explicit evidence in field theory~\cite{Osborn:1979,Sen:1994,Vafa:1994} and through the relation between string theory and the $N=4$ theory~\cite{Schwarz:1993}. In this paper we consider the computation of certain terms in the partition function, referred to as the BPS terms, that are independent of the coupling constant and can therefore be reliably computed in the weak coupling limit. The contribution from topologically trivial gauge field configurations, however, is not accessible by direct computations, due to the presence of zero modes, but can be obtained using $S$-duality. The self-consistency of the full solution, in addition, provides some insight into the underlying structure of $S$-duality of the BPS terms.

In order to retain maximal supersymmetry without having to invoke a topological twisting we consider the $N=4$ theory defined on a flat Euclidean four-torus
\beq
M = T^4 = \R^4 / \Gamma ,
\eeq
where $\Gamma \subset \R^4$ is a rank four lattice. We take the local gauge group $G$ of the theory to be of adjoint type, i.e. 
\beq
G =  \hat{G} / \cC ,
\eeq
where $\hat{G}$ is the simply connected universal covering group of $G$, and $\cC$ is the abelian center subgroup of $\hat{G}$. The order of $\cC$ is denoted $| \cC |$. We will specialize to the case when $\hat{G} \simeq \SU (n)$ for some prime number $n$ so that $\cC \simeq \Z / n \Z$ and $| \cC | = n$.

The global $R$-symmetry group of the theory is 
\beq
\cR \simeq \SU (4) \simeq \Spin (6) .
\eeq
We couple the corresponding currents to a non-dynamical flat connection $B$ on a topologically trivial principal $\cR$-bundle over $M$. Up to simultaneous conjugation by elements of $\cR$, such a connection is equivalent to a homomorphism (given by the commuting holonomies) $\pi_1 (M) \rightarrow \cT$, where $\cT$ is an arbitrary maximal torus subgroup (a Cartan torus) of $\cR$. Furthermore, $\pi_1 (M) \simeq H_1 (M, \Z)$ for $M = T^4$, so the flat connection can be identified with an element 
\beq
B \in H^1 (M, \cT) .
\eeq 
Note that the coupling to a generic non-vanishing $\cR$-symmetry connection breaks all supersymmetries of the theory. We will mainly be interested in the connections that leave one of the supersymmetries unbroken.

In the next section we discuss the decomposition of the theory into sectors labelled by the 't~Hooft flux $v \in H^2(M,\cC)$ which may be studied separately. The corresponding partition functions are denoted by $Z_v (\tau | \Gamma, B)$, and can be viewed as components of a vector in a linear space. As the notation indicates they depend on the coupling $\tau$ as well as the geometry of the manifold $M$ and the background $\cR$-symmetry connection. We also describe the action of S-duality on the components $Z_v (\tau | \Gamma, B)$. In section three we compute the partition functions $Z_v (\tau | \Gamma, B)$ in the $\tau \rightarrow i \infty$ weak coupling limit for all non-trivial 't~Hooft fluxes $v \neq 0$. (Because of its subtle infrared behaviour, the partition function $Z_0 (\tau | \Gamma, B)$ for trivial 't~Hooft flux does not appear to directly calculable even at weak coupling, as mentioned above.) In section four we will then argue that when the partition functions $Z_v(\tau|\Gamma,B)$ (including the $Z_0 (\tau | \Gamma, B)$ component) are expanded in a power series in $B$ around the loci where one of the supersymmetries is unbroken, the constant and linear terms are in fact independent of $\tau$. Thus, the result at weak coupling can be used to compute these terms for $v \neq 0$ and the result may be extrapolated to the strong coupling regime. Furthermore, we will show that the a priori overdetermined system of equations imposed by $S$-duality in fact does admit a unique solution for the constant and linear terms of $Z_0 (\tau | \Gamma, B)$. The solution is conveniently formulated in terms of the generalization to arbitrary 't~Hooft flux of a certain set of shifts of allowed momenta appearing in the computation of the partition function $Z_v (\tau | \Gamma, B)$ for rank one $v$ at weak coupling.

\section{The 't Hooft flux and S-duality}
In addition to the connection $A$ with field strength $F$ on the gauge bundle $P$, which is a principal $G$-bundle over $M$, the $N=4$ theory contains scalar and spinor fields $\Phi$ and $\Psi$. The transformation properties of the various fields can be described by
\bea
F & \in & \Omega^2 (M, {\rm ad} (P) ) \cr
\Phi & \in & \Omega^0 (M, {\rm ad} (P) \otimes V_{\bf 6}) \cr
\Psi & \in & \Omega^0 (M, {\rm ad} (P) \otimes V_{\bf 4} \otimes S) ,
\eea
where ${\rm ad \,} (P)$ is the vector bundle associated to $P$ via the adjoint representation of $G$, the vector bundles $V_{\bf 4}$ and $V_{\bf 6}$ are associated to the $R$-symmetry bundle via the fundamental representation ${\bf 4}$ and the anti-symmetric rank two tensor representation ${\bf 6}$ of $\cR$ respectively, and the vector bundle $S$ is the trivial spin bundle. The theory is completely described by the action
\bea
S & = & \frac{{\rm Im \,}  \tau}{4 \pi} \int_M {\rm Tr} \left( F \wedge *F + (D + iB) \Phi \wedge * (D + iB) \Phi \right. \cr 
& & \left. + {\rm Vol}_M \bar{\Psi} (\slash \!\!\!\! D + i \slash \!\!\!\! B) \Psi  + \ldots \right)+ \frac{i \, {\rm Re \,} \tau}{4 \pi} \int_M {\rm Tr} \left( F \wedge F \right) ,
\eea
where $D$ is the covariant derivative corresponding to the gauge connection $A$, and $\tau$ is the complex coupling constant 
\beq
\tau = \frac{\theta}{2 \pi} + \frac{4 \pi i}{g^2} 
\eeq
taking its values in the complex upper half-plane. As usual, $g$ and $\theta$ denote the coupling constant and the theta-angle respectively. The ellipsis in $S$ denote interaction terms of type $\Phi \Psi \Psi$ and $\Phi \Phi \Phi \Phi$ which are suppressed in the weak coupling limit. The normalization is appropriate for $\hat{G} = \SU (n)$ with ${\rm Tr}$ denoting the trace in the fundamental representation\footnote{With this normalization the instanton number introduced below can take arbitrary integral value for $\hat{G} =\SU(n)$.}.

\subsection{Topology of the gauge bundle}
The isomorphism class of the gauge bundle $P$ over the four-dimensional base manifold $M$ is completely determined by the 't~Hooft flux
\beq
v \in H^2 (M, \cC) \simeq H^2 (M, \Z) \otimes \cC
\eeq
and the fractional instanton number (or second Chern class)
\beq
k \in H^4(M,\R) \simeq H^0(M,\R) .
\eeq
However, these characteristic classes are not completely independent: For a given value of $v$ the allowed values of $k$ are given by the relation
\beq
\label{eqn:tHooftFluxInstantonRelation}
k - \frac{1}{2} v \cdot v \in H^4(M,\Z) \subset H^4(M,\R) .
\eeq

In the condition above we have introduced the product 
$v \cdot v^\prime \in H^4 (M, \R / \Z)$ of two elements $v, v^\prime \in H^2 (M, \cC)$ which is defined as the composition of the cup product 
\beq
\cup : H^2 (M, \Z) \times H^2 (M, \Z) \rightarrow H^4 (M, \Z)
\eeq
with the non-degenerate and symmetric pairing 
\beq
\cC \times \cC \rightarrow \R / \Z ,
\eeq
which follows from the interpretation
\beq
\cC \simeq \Gamma^{\rm weight} / \Gamma^{\rm root}
\eeq
of $\cC$ in terms of the dual weight and root lattices $\Gamma^{\rm weight}$ and $\Gamma^{\rm root}$ of $\hat{G}$. For the case $\hat{G}=\SU(n)$ we can identify $\cC = \Z/n\Z$, and the pairing of two elements is computed by first lifting to $\Z$, multiplying, dividing by $n$ and finally reducing modulo $\Z$, so that
\beq
\label{eqn:tHooftFluxProduct}
v \cdot v^\prime \in \frac{1}{n} \Z / \Z \subset \R / \Z \simeq H^0 (M,\R / \Z) \simeq H^4 (M,\R / \Z) ,
\eeq
Furthermore, since the manifold $M = T^4$ is spin, the product $v \cdot v$ is divisible by two in a canonical way and the resulting class, the lift of which appears in the condition (\ref{eqn:tHooftFluxInstantonRelation}), is called the Pfaffian of $v$
\beq
{\rm Pf}(v) = \frac{1}{2} v \cdot v  \in H^4 (M, \R / \Z) .
\eeq
We will use the notation ${\rm Pf}(v)$ interchangeably for both the cohomology class $\frac{1}{2} v \cdot v$ and its integral over $M$, exploiting the isomorphisms displayed in (\ref{eqn:tHooftFluxProduct}).

\subsection{The mapping class group}
At this point it is appropriate to recall the following facts about $H^2 (M, \cC)$ for $M = T^4 = \R^4 / \Gamma$ and $\cC \simeq \Z / n \Z$: By a choice of basis $e^1, e^2, e^3, e^4$ of $H^1 (M, \Z)$, the mapping class group of $M$ may be identified with the arithmetic group $\SL(4,\Z)$ and acts on $v \in H^2 (M, \cC)$ via its modulo $n$ reduction $\SL(4,\cC)$. The Pfaffian $\frac{1}{2}v\cdot v$ is one of the two\footnote{The second invariant is the greatest common divisor modulo $n$, ${\rm gcd}(v_{ij},n)$, of $n$ and the components $v_{ij}$ of $v$. However this quantity is not independent of Pf$(v)$ when $n$ is prime and ${\rm Pf}(v) \neq 0$. To be more specific, ${\rm gcd}(v_{ij},n)=0$ for $v=0$ and ${\rm gcd}(v_{ij},n)=1$ for $v \neq 0$.} $\SL(4,\cC)$-invariants that can be constructed from $v$, and the condition relating $v$ and $k$ is thus also $\SL(4,\cC)$-invariant.

We can furthermore consider the characterization of the $| \cC |^6 = n^6$ different possible values of the 't~Hooft flux $v$ by their orbits under $\SL(4,\cC)$, which is of order
\beq
\left| \SL(4,\cC) \right| = n^6 (n^2 - 1) (n^3 - 1) (n^4 - 1) .
\eeq
For this purpose it is also useful to note the orders of some of its subgroups:
\beq
\left| \SL(2,\cC) \right| = n (n^2 - 1)
\eeq
and
\beq
\left| \Sp(4,\cC) \right| = n^4 (n^2 - 1) (n^4 - 1) .
\eeq
The complete list of $\SL(4,\cC)$ orbits on $H^2 (M, \cC)$ for $n$ prime,  with representative elements, stabilizing subgroups of these and cardinalities (given as the ratio between the orders of $\SL(4,\cC)$ and the stabilizers) is given in table \ref{tab:SL4COrbits}.
\begin{table}[!ht]
{\small
\beq
\begin{array}{llll}
{\rm Orbit} & {\rm Representative} & {\rm Stabilizer} & {\rm Cardinality} \cr
\hline
v = 0 & 0 & \SL(4,\cC) & 1 \cr
\frac{1}{2} v \cdot v = 0, v \neq 0 & e^1 \cup e^2 & \SL(2,\cC) \times \SL(2,\cC) \times \cC^4 & n^5 + n^3 - n^2 - 1 \cr
\frac{1}{2} v \cdot v = \frac{1}{n} & e^1 \cup e^2 + e^3 \cup e^4 & \Sp(4,\cC) & n^5 - n^2 \cr
\frac{1}{2} v \cdot v = \frac{2}{n} & e^1 \cup e^2 + 2 e^3 \cup e^4 & \Sp(4,\cC) & n^5 - n^2 \cr
\ldots & \ldots & \ldots & \ldots \cr
\frac{1}{2} v \cdot v = \frac{n - 1}{n} & e^1 \cup e^2 + (n - 1) e^3 \cup e^4 & \Sp(4,\cC) & n^5 - n^2 \cr
\hline
& & & n^6
\end{array} \nonumber 
\eeq
\caption{Orbits of $\SL(4,\cC)$ on $H^2(M,\cC)$ according to the value of $\frac{1}{2}v \cdot v$.}
\label{tab:SL4COrbits}
}
\end{table}

In the considerations of the present paper we will often distinguish between three classes of $\SL(4,\cC)$-orbits according to the rank of the 't~Hooft flux. In particular, all $n-1$ orbits with $\frac{1}{2}v \cdot v \neq 0$ have similar properties and are conveniently grouped together. We will therefore refer to such $v$ as rank two 't~Hooft fluxes, while $\frac{1}{2} v \cdot v = 0, v \neq 0$ and $v=0$ are referred to respectively as rank one and rank zero fluxes.

\subsection{S-duality}
In addition to the gauge group $G$ of the $N=4$ theory, the $S$-duality conjecture involves the (Langlands) dual group $G^{\vee}$. In the case presently under consideration we have $G \simeq \SU(n)/\cC$, the dual of which is $G^{\vee} \simeq \SU(n)$. In a path integral representation of the partition function, the decomposition of the theory into sectors labelled by the 't~Hooft flux is manifest. In particular, since the 't~Hooft flux of the $\SU(n)$ theory is necessarily trivial, we have
\beq
Z_G(\tau | \Gamma,B) = \sum_v Z_v(\tau | \Gamma,B)
\eeq
and
\beq
Z_{G^{\vee}}(\tau | \Gamma,B) = |\cC|^3 Z_0(\tau|\Gamma,B) ,
\eeq
where the factor $|\cC|^3$ compensates for the fact that the volume of the group of $\SU(n)$ gauge transformations differs from that of the group of $\SU(n) / \cC$ transformations~\cite{Vafa:1994}. The components $Z_v(\tau | \Gamma,B)$, defined as the contributions to the partition function of the theory with gauge group $G$ from gauge bundles with a particular 't~Hooft flux $v$, are given by
\beq
Z_v(\tau | \Gamma,B) = \sum_k  \int \cD A \cD \Phi \cD \Psi \exp (-S) ,
\eeq
where the sum is over all instanton numbers $k$ satisfying (\ref{eqn:tHooftFluxInstantonRelation}), i.e. over all isomorphism classes of bundles with 't~Hooft flux $v$. 

The original $S$-duality conjecture amounts to the statement that an $\SL(2,\Z)$ transformation of the complex coupling constant, (possibly) combined with the exchange of the gauge group $G \simeq \SU(n)/\cC$ for its dual $G^{\vee} \simeq \SU(n)$, maps the original theory to an equivalent one. In particular, the $\Z_2$ duality of Montonen and Olive can be expressed as
\beq
\label{eqn:Montonen_Olive}
Z_{G^{\vee}}(-\frac{1}{\tau}|\Gamma,B) = Z_G(\tau|\Gamma,B) .
\eeq

Given the decomposition of the theory according to the 't~Hooft flux $v$, the conjecture can be extended~\cite{Vafa:1994} to a unitary linear relationship between the components $Z_v(\tau | \Gamma,B)$ for values of $\tau$ related by
\bea
\tau \mapsto \frac{a \tau + b}{c \tau + d} & , & \left( \begin{matrix}a & b \cr c & d \end{matrix} \right) \in \SL(2,\Z) .
\eea
This group of transformations is generated by the $S$-transformation
\beq
\label{eqn:STransformation}
Z_v (- 1 / \tau | \Gamma, B) = | \cC |^{-3} \sum_{v^\prime} \exp \left( 2 \pi i \int_M v \cdot v^\prime \right) Z_{v^\prime} (\tau | \Gamma, B)
\eeq
and the $T$-transformation
\beq
\label{eqn:TTransformation}
Z_v (\tau + 1Ê| \Gamma, B) = \exp \left( - 2 \pi i \int_M \frac{1}{2} v \cdot v \right) Z_v (\tau | \Gamma, B) ,
\eeq
subject to the relations
\beq
S^2 = (S T)^3 = 1 .
\eeq

We note that the $T$-transformation is manifestly satisfied for each component $Z_v$ due to the relation (\ref{eqn:tHooftFluxInstantonRelation}) and the fact that $\frac{1}{8\pi^2}{\rm Tr} (F \wedge F)$ is a representative in de Rham cohomology of the instanton number $k$. The $S$-transformation, however, is a non-trivial condition on the components $Z_v$. To conclude this section, we also note that the relation (\ref{eqn:Montonen_Olive}) follows from (\ref{eqn:STransformation})~\cite{Vafa:1994}.

\section{The weak coupling limit}
We now proceed to consider the $N=4$ theory in the limit $\tau \to i\infty$ of weak gauge coupling, and in particular attempt to compute the partition functions $Z_v (\tau | \Gamma, B)$. To achieve this, we first note that all the real terms in $S$ are positive semi-definite, and for a given instanton number $k$ we have
\beq
{\rm Re \,} S \geq \frac{1}{g^2} \int_M {\rm Tr} (F \wedge * F) \geq \frac{8 \pi^2}{g^2} \left| \int_M k \right| ,
\eeq
so contributions to $Z_v (\tau | \Gamma, B)$ from gauge bundles of non-zero $k$ are exponentially suppressed in the limit $g \rightarrow 0$. 

Depending on the $\SL(4,\cC)$-orbit of the 't~Hooft flux $v \in H^2 (M, \cC)$, we may now distinguish between the following cases:
\begin{itemize}
\item
For $v$ such that $\frac{1}{2} v \cdot v \neq 0$ we have $k \neq 0$ for all gauge bundles, and thus
\beq
Z_v (\tau | \Gamma, B) \rightarrow 0
\eeq
in the weak coupling limit. 
\item
For $v \neq 0$ but such that $\frac{1}{2} v \cdot v = 0$, there is a unique gauge bundle with $k = 0$. Furthermore, this bundle admits a unique flat connection $\cA$. The weak coupling limit of the partition function $Z_v (\tau | \Gamma, B)$ can then be computed in a one-loop approximation around such a background, as we will describe in detail below.
\item
For $v = 0$, there is again a unique gauge bundle with $k = 0$, but this bundle admits a moduli space of flat connections of real dimension $4(n - 1)$. Because of problems with zero-modes, we cannot perform a direct computation of the corresponding partition function $Z_0 (\tau | \Gamma, B)$, even in the weak coupling limit. (In the last section, however, we will be able to gain some information about it via $S$-duality.)
\end{itemize}
The above results, valid for $n$ prime, can be obtained using a parametrization of the moduli space of flat connections of gauge bundles over $T^4$ by conjugacy classes of quadruples of almost commuting elements in $\SU(n)$. This method is completely analogous to that previously used to study flat connections over $T^3$~\cite{Witten:1998,Keurentjes:1999a,Keurentjes:1999b,Kac:2000,Borel:1999,Witten:2000}.

\subsection{Non-trivial 't~Hooft flux}
As we saw above the partition function $Z_v(\tau|\Gamma,B)$ vanishes in the weak coupling limit for 't~Hooft flux $\frac{1}{2}v \cdot v \neq 0$. We therefore concentrate on the remaining non-trivial orbit, i.e. $v \neq 0$ but $\frac{1}{2} v \cdot v = 0$. An arbitrary connection $A$ on the gauge bundle with $k = 0$ can then be written in terms of a quantum fluctuation 
\beq
a \in \Omega^1 (M, {\rm ad} (P))
\eeq 
around the unique flat connection $\cA$ as
\beq
A =  \cA + g a .
\eeq
We also rescale the scalar and spinor fields $\Phi$ and $\Psi$ according to
\bea
\Phi & = & g \phi \cr
\Psi & = & g \psi .
\eea
As we will see shortly, the factors of the coupling constant $g$ are inserted to make the fields $a$, $\phi$ and $\psi$ canonically normalized. For the purpose of gauge fixing, we also introduce a complex fermionic ghost field
\beq
\omega \in \Omega^0 (M, {\rm ad} (P)) .
\eeq

In generalized Feynman background gauge with real parameter $\xi$, the action in terms of these variables can then be expanded as
\bea
S & = & \int_M {\rm Tr} \Bigl( * a \wedge (* \cD * \cD + \frac{1}{\xi} \cD * \cD *) a - * \phi \wedge * (\cD + iB) * (\cD + iB) \phi \cr
& & + {\rm Vol}_M \bar{\psi} (\slash \!\!\!\! \cD + i\slash \!\!\!\! B) \psi  + * \bar{\omega} \wedge * \cD * \cD \omega \Bigr) + \cO (g) ,
\eea
where $\cD$ is the covariant derivative corresponding to the flat background gauge connection $\cA$. We then have
\beq
Z_v (\tau | \Gamma, B) = Z_v^{\rm one-loop} (\Gamma, B) + \cO (g^2) ,
\eeq
with the $\tau$-independent one-loop contribution given by a Gaussian functional integration as
\beq
\label{eqn:OneLoopPartitionFunction}
 Z_v^{\rm one-loop} (\Gamma, B) = \frac{\det (* \cD * \cD) \det (\slash \!\!\!\! \cD + i\slash \!\!\!\! B)}{\det^{1/2} (* \cD * \cD + \frac{1}{\xi} \cD * \cD *) \det^{1/2} (* (\cD + iB) * (\cD + iB))} .
\eeq

To proceed further, we need to compute the spectrum of the various differential operators appearing in this expression. We begin by considering the holonomies
\beq
U_{\gamma} = \exp \left( i \int_\gamma \cA \right)
\eeq
of the background connection $\cA$ around one-cycles $\gamma \in H_1 (M, \Z)$. Their adjoint actions on the Lie algebra $\mathfrak{g}=\su(n)$ of $G$ commute and may thus be simultaneously diagonalized. Identifying once again $\cC=\Z/n\Z$ we can thus introduce a basis $\{T_{\rho}\}$ of $\mathfrak{g}$, where $\rho \in H^1(M,\cC)$, satisfying
\beq 
\label{eqn:HolonomyEigenvalue}
U_{\gamma} T_\rho U_{\gamma}^{-1} = \exp \left( \frac{2 \pi i}{n} \int_\gamma \rho \right) T_\rho ,
\eeq
for all $\gamma \in H_1 (M, \Z)$. The eigenvalues under conjugation are complex roots of unity by virtue of the finite order of the holonomies. For a given value of $v$ the corresponding $\rho$ in (\ref{eqn:HolonomyEigenvalue}) constitute the set 
\beq
\label{eqn:RhoEigenvalues}
S_v = \left\{ \rho \in H^1 (M, \cC) | \rho \neq 0, \rho \cdot v = 0 \right\} ,
\eeq
with cardinality $n^2-1$ in agreement with the dimension of $\mathfrak{g}$. Once again, the product $\rho \cdot v$ is defined as the composition of the cup product and the pairing on $\cC$.

The above result can be derived by considering a particular element of the rank one orbit, e.g. $v = e^1 \cup e^2$. We then have 
\beq
S_v = \left\{ (\rho_1,\rho_2,0,0) | (\rho_1,\rho_2) \neq (0,0) \right\} ,
\eeq
where the components $\rho_i$ refer to the standard basis $e^1,e^2,e^3,e^4$, which is an equivalent presentation of (\ref{eqn:RhoEigenvalues}). (The analogous result on $T^3$ was derived in~\cite{LindmanHornlund:2008}.) Using the $\SL(4,\cC)$-covariance of the condition $\rho \cdot v = 0$ we then obtain the result (\ref{eqn:RhoEigenvalues}) for all $v$ in the $\frac{1}{2}v \cdot v = 0 \,,\, v \neq 0$ orbit.

Furthermore, for $\rho \in H^1 (M, \cC)$, we introduce the set 
\beq \label{eqn:PRho}
P_\rho = \left\{ p \in H^1 (M, \frac{1}{n} \Z) \Bigl| [n p] = \rho \right\} ,
\eeq
where we let $[n p]$ denote reduction modulo $n$ of the integral class $np$. The space of sections of the vector bundle ${\rm ad} (P)$ then has a basis $\{ u_p \}$, where $p$ runs over the $v$-dependent set
\beq
\label{eqn:PSet}
P_v = \underset{\rho \in S_v}{\bigcup} P_\rho .
\eeq
The basis elements $u_p$ are simultaneous eigensections of the covariant directional derivatives on $M$, satisfying
\beq
\cD_{\mu} u_p = 2 \pi i p_{\mu} \, u_p ,
\eeq
where $\mu=1,2,3,4$ refers to the basis $e^1,e^2,e^3,e^4$, and the $p \in P_v$ consequently constitute the allowed momenta on $M$ for gauge bundles of 't~Hooft flux $v$. Once again, this result can be obtained by extending the construction of~\cite{LindmanHornlund:2008} to a four dimensional torus. This determines the spectra of the operators $* \cD * \cD$ and $* \cD * \cD + \frac{1}{\xi} \cD * \cD *$ that are constructed only from $\cD$ (and the metric on $M$).

To determine also the spectra of the operators involving the $R$-symmetry connection $B \in H^1 (M, \cT)$, it is convenient to describe an element of the Cartan torus $\cT \subset \cR$ by the phases $\exp (2 \pi i t^1)$, $\exp (2 \pi i t^2)$, $\exp (2 \pi i t^3)$ and $\exp (2 \pi i t^4)$ by which it acts on the weight spaces $V_{w_1}$, $V_{w_2}$, $V_{w_3}$, and $V_{w_4}$ of the module 
\bea
V_{\bf 4} = \underset{\,\, w_i}{\bigoplus} \, V_{w_i}& , & i=1,2,3,4
\eea
of the fundamental ${\bf 4}$ representation of $\cR$. The weights obey $w_1 + w_2 + w_3 + w_4 = 0$ so we can identify
\beq
\cT = \left\{ (t^1, t^2, t^3, t^4) \in (\R / \Z)^4 | t^1 + t^2 + t^3 + t^4 = 0 \right\} .
\eeq
The $R$-symmetry connection then amounts to a quartet $B = (B^1, \ldots, B^4)$ of elements of the cohomology group $H^1 (M, \R / \Z)$ subject to the relation
\beq
B^1 + B^2 + B^3 + B^4 = 0 .
\eeq
We take the liberty to denote by $B^i$ also the lift of the corresponding elements of $H^1(M,\R/\Z)$ to $H^1(M,\R)$, which is performed in such a way as to preserve the property $\sum_i B^i = 0$. Hopefully, this will not cause confusion as it should be clear from the context in which set the components of the $B^i$ are valued. 

In the differential operators we are presently considering, the $\cR$-symmetry connection enters through minimal coupling in the covariant derivate $\cD+iB$ acting on the various fields in the $N=4$ theory. In this context $B$ should be considered as the connection 1-form of the principal $\cR$-bundle over $M$. Since $B$ is flat it is actually valued in the Cartan subalgebra of the Lie algebra of $\cR$, and therefore related to the representation in the previous paragraph by the exponential map. (Connections $B$ that exponentiate to the same element in $H^1(M,\cT)$ are related by global bundle automorphisms (gauge transformations) of the $\cR$-bundle.) Consequently, the action of $B$ on a vector $v_i \in V_{w_i}$ is given by
\beq
B v_i =  2 \pi  B^i v_i ,
\eeq
where there is no sum on $i$. The extension to arbitrary representations of $\cR$ follows from the tensor product construction.

\subsection{The one-loop partition function}
We are now ready to describe the spectra of the differential operators appearing in (\ref{eqn:OneLoopPartitionFunction}) explicitly, recalling that the spinor and scalar fields $\psi$ and $\phi$ transform in the ${\bf 4}$ and ${\bf 6}$ representations respectively, and therefore have $\cR$-symmetry weights $w_1,w_2,w_3,w_4$ and $w_1+w_2,w_1+w_3,w_1+w_4,w_2+w_3,w_2+w_4,w_3+w_4$. For each allowed momentum $p$ in the set (\ref{eqn:PSet}) we then have (up to constant factors that can be absorbed by rescaling the fields when performing the Gaussian path integrals):
\begin{itemize}
\item
For the ghost Laplacian $* \cD * \cD$ a non-degenerate eigenvalue 
\beq
| p |^2 ,
\eeq
where the norm is given by the metric on $M = T^4 = \R^4 / \Gamma$. 
\item
For the gauge field operator $* \cD * \cD + \frac{1}{\xi} \cD * \cD *$ a non-degenerate eigenvalue 
\beq
\xi^{-1} | pÊ|^2
\eeq
and a triply degenerate eigenvalue 
\beq
| pÊ|^2, | pÊ|^2, | pÊ|^2 .
\eeq
\item
For the Dirac operator $\slash \!\!\!\! \cD + i\slash \!\!\!\!  B$ the non-degenerate eigenvalues 
\beq
| p + B^1 |, | p - B^1 |, | p + B^2 |, | p - B^2 |, | p + B^3 |, | p - B^3 |, | p + B^4 |, | p - B^4 | .
\eeq
\item
For the scalar Laplacian $* (\cD + iB) * (\cD + iB)$ the non-degenerate eigenvalues
\bea
&| p + B^1 + B^2 |^2, | p + B^1 + B^3 |^2, | p + B^1 + B^4 |^2 \, & \cr
&| p + B^2 + B^3 |^2, | p + B^2 + B^4 |^2, | p + B^3 + B^4 |^2 . & 
\eea
\end{itemize}
Putting everything together, we get the one-loop partition function
\beq
\label{eqn:ZvOneLoop}
Z_v^{\rm one-loop} (\Gamma, B) = \prod_{\rho \in S_v} \prod_{p \in P_{\rho}} \frac{1}{|p|^2} \prod_{i=1}^{4} |p+B^i| |p-B^i| \prod_{\substack{j,k = 1\\ j<k}}^{4} \frac{1}{|p+B^j+B^k|} ,
\eeq
where we have normalized by dividing with $Z_v^{\rm one-loop}(\Gamma,0)$ to avoid constant divergent factors. We have also split the product over momenta according to the elements $\rho$ of the set $S_v$, which will prove convenient in the remainder of the considerations of this paper. 

However, one could still worry about the ultraviolet convergence of the infinite product over momenta $p \in P_{\rho}$. To investigate potential divergences in the expression above we compute its logarithm and perform a Taylor expansion to isolate the dependence on powers of $|p|$. Exponentiating to recover the one-loop partition function we obtain
\beq
Z_v^{\rm one-loop} (\Gamma, B) = \exp \Bigl( \sum_{n = 1}^\infty \sum_{\rho \in S_v} \sum_{p \in P_\rho} |p|^{-2 n} W_v^{(2 n)} (\Gamma, \hat{p}, B) \Bigr) ,
\eeq
where $\hat{p} = p / |p| \in S^3$ and the coefficient $W_v^{(2 n)} (\Gamma, \hat{p}, B)$ is of order $2 n$ in $B^1$, $B^2$, $B^3$, $B^4$. The terms with $n = 1$ and $n = 2$ are potentially quadratically and logarithmically divergent respectively. Fortunately one finds that the $n = 1$ term vanishes identically
\beq
W_v^{(2)} (\Gamma, \hat{p}, B) = 0 .
\eeq
Furthermore, the coefficient of the $n = 2$ term is
\bea
W_v^{(4)} (\Gamma, \hat{p}, B) & = & 2 \left(B^1 \! \cdot \! B^2 B^3 \! \cdot \! B^4 + B^2 \! \cdot \! B^3 B^1 \! \cdot \! B^4 + B^3 \! \cdot \! B^1 B^2 \! \cdot \! B^4) \right. \cr
& & - 8 \left( \hat{p} \cdot \! B^1 \hat{p} \cdot \! B^2 B^3 \! \cdot \! B^4 + \hat{p} \cdot \! B^1 \hat{p} \cdot \! B^3 B^2 \! \cdot \! B^4 + \hat{p} \cdot \! B^1 \hat{p} \cdot \! B^4 B^2 \! \cdot \! B^3 \right. \cr
& & \left. + \hat{p} \cdot \! B^2 \hat{p} \cdot \! B^3 B^1 \! \cdot \! B^4 + \hat{p} \cdot \! B^2 \hat{p} \cdot \! B^4 B^1 \! \cdot \! B^3 + \hat{p} \cdot \! B^3 \hat{p} \cdot \! B^4 B^1 \! \cdot \! B^2 \right) \cr
& & + 48 \hat{p} \cdot \! B^1 \hat{p} \cdot \! B^2 \hat{p} \cdot \! B^3 \hat{p} \cdot \! B^4 ,
\eea
where the raised dot denotes the inner product given by the metric on $M$. In the ultraviolet, the sum over $p\in P_{\rho}$ can be compared with an integral over $\R^4$ (with the standard metric), which can be performed in spherical coordinates:
\beq
\sum_{p \in P_\rho} \approx \int_{\R^4} d^4 p = \int_0^\infty | p |^3 d | p| \int_{S^3} d^3 \hat{p} .
\eeq
For the case $\Gamma=\Z^4$ the coefficient of the logarithmically diverging term is then
\bea
\label{eqn:CoefficientLogDivergence}
\int_{S^{d - 1}} d^{d - 1} \hat{p} \, W_v^{(4)} (\Z^4, \hat{p}, B) & = & (B^1 \! \cdot \! B^2 B^3 \! \cdot \! B^4 + B^2 \! \cdot \! B^3 B^1 \! \cdot \! B^4 + B^3 \! \cdot \! B^1 B^2 \! \cdot \! B^4) \cr
& & \times \frac{4 \pi^{d / 2} (d - 2) (d - 4)}{\Gamma (d / 2) d (d + 2)} ,
\eea
which vanishes for $d = 4$. For a general lattice $\Gamma$, a linear coordinate transformation in momentum space is required when evaluating the integral to obtain the standard metric on $\R^4$. The Jacobian of this transformation introduces a multiplicative constant in the integral. Consequently, the coefficient of the logarithmically diverging term is proportional to (\ref{eqn:CoefficientLogDivergence}) and vanishes for an arbitrary flat torus $M=T^4$. A more precise definition of $Z_v^{\rm one-loop} (\Gamma, B)$ would thus amount to introducing a temporary cutoff $\Lambda$, which is eventually taken to infinity:
\beq
Z_v^{\rm one-loop} (\Gamma, B) = \lim_{\Lambda \rightarrow \infty} \prod_{\rho \in S_v} \prod_{\substack{p \in P_{\rho} \\ |p| < \Lambda}} \frac{1}{|p|^2} \prod_{i=1}^{4} |p+B^i| |p-B^i| \prod_{\substack{j,k = 1\\ j<k}}^{4} \frac{1}{|p+B^j+B^k|} ,
\eeq

These good convergence properties seem to be unrelated to the usual more familiar ultraviolet finiteness properties of $N = 4$ supersymmetric Yang-Mills theory. One the one hand, e.g. for a generic supersymmetric (but not maximally supersymmetric) Yang-Mills theory with vanishing one-loop beta-function, already the coefficient of the $n = 1$ term in the sum above would not vanish even after integrating over $d^3 \hat{p}$. But on the other hand, infinities are known to arise also in $N = 4$ theories, leading e.g. to anomalous scaling dimensions.

\section{The BPS terms}
We have thus far computed the components $Z_v (\tau | \Gamma, B)$ in the weak coupling limit $\tau \rightarrow i \infty$ for $v \neq 0$. But for finite $\tau$, (perturbative and non-perturbative) corrections must in general be taken into account. Also we have not been able to perform any computations for $v = 0$ due to the presence of zero modes. 

To find quantities that are in fact independent of $\tau$, we expand the partition function around some $B = (B^1, B^2, B^3, B^4)$ that leaves one of the weights of the fundamental representation ${\bf 4}$ of $\cR$ invariant, i.e. which acts trivially on the corresponding weight space. We may e.g. take $B^4 = 0$, which leaves $V_{w_4}$ invariant. At the Lie algebra level the equivalent statement is that $B$ annihilates all elements of that weight space, thus leaving the corresponding supersymmetry unbroken while breaking the supersymmetry generators associated to the weights $w_1$, $w_2$ and $w_3$.

We let $\delta B = (0,0,0, \delta B^4)$ denote the deviation from $B$. Here, we consider $B^1,B^2,B^3$ to be exact, absorbing any perturbation of these components into their definition. Once again we use the notation $\delta B$ also for the minimal lift (ensuring that perturbation theory remains meaningful) from $H^1(M, \R / \Z)$ to $H^1(M, \R)$ of the deviation. We can then expand $Z_v (\tau | \Gamma, B + \delta B)$ in a power series in $\delta B$ to obtain
\beq
Z_v (\tau | \Gamma, B + \delta B) = Z_v^{(0)} (\Gamma, B) + Z_v^{(1)} (\Gamma, B | \delta B) + \cO ((\delta B)^2) ,
\eeq
where $Z_v^{(0)} (\Gamma, B)$ and $Z_v^{(1)} (\Gamma, B | \delta B)$ denote respectively the terms of order zero and one in $\delta B$. We will now show that these terms (in contrast to the omitted terms of second or higher order in $\delta B$) are independent of $\tau$, so they can be computed in a one-loop approximation at weak coupling and the result reliably extrapolated to arbitrary values of the coupling.

To show that $Z_v^{(0)} (\Gamma, B)$ and $Z_v^{(1)} (\Gamma,B|\delta B)$ indeed do not depend on the coupling it is convenient to consider a Hamiltonian formulation (in Minkowski signature): The partition function is then given by
\beq
\label{eqn:HamiltonPartitionFunction}
Z_v (\tau | \Gamma, B) = {\rm Tr}_{\cH_v} \left( (-1)^F \exp ( - i t \hat{H} + i {\bf x} \cdot \hat{P} + i B_0) \right) ,
\eeq
where $\cH_v$ is the Hilbert space of the sector of the theory with gauge bundles of 't~Hooft flux $v$. Furthermore, $\hat{H}$ and $\hat{P}$ are the Hamiltonian and the momentum operators, $B_0$ is the 'time' component of the $\cR$-symmetry connection and $F$ is the operator that measure the fermion number. The supersymmetry generators can be seen as two sets of fermionic creation and annihilation operators transforming in the representations ${\bf 4}$ and $\overline{\bf 4}$ respectively under $\cR \simeq \Spin (6) \simeq \SU (4)$. But because of the spatial components of the $R$-symmetry connection $B=(B^1, B^2, B^3, 0)$, around which we expand, only the two creation operators of weight $w_4$ and the two annihilation operators of weight $- w_4$ are generically unbroken. Consequently, all states in the Hilbert space fall into $N=1$ multiplets which, depending on the value of the four-momentum $(E, {\bf p})$, are divided into three classes: 
\begin{itemize}
\item
Vacuum states, which have $E^2 = {\bf p}^2 = 0$, are invariant under all supersymmetries and transform trivially under $\cR$. They thus contribute only to the constant term $Z_v^{(0)} (\Gamma, B)$ in the partition function.
\item
Light-like states, for which $E^2 = {\bf p}^2 > 0$, are half-BPS in the sense that half of the supersymmetries (i.e. one set of creation and annihilation operators) act trivially. Because of the insertion $(-1)^F \exp (i B_0)$ there is thus a factor
\beq
1 - \exp (2 \pi i \delta B_0^4) = \cO (\delta B) , 
\eeq
where we recall that $B^4=0$, in the contribution to the partition function. Consequently, these states give a contribution to $Z_v^{(1)} (\Gamma, B|\delta B)$.  
\item
Generic states, for which $E^2 > {\bf p}^2 \geq 0$, instead give a contribution proportional to
\beq
1 - 2 \exp (2 \pi i \delta B_0^4) + \exp (4 \pi i \delta B_0^4) = \cO ((\delta B)^2)
\eeq
to the partition function, and thus do not contribute to either of the terms $Z_v^{(0)} (\Gamma, B)$ or $Z_v^{(1)} (\Gamma, B, |\delta B)$. 
\end{itemize}

It should be noted that while the half-BPS states give contributions also to $\cO ((\delta B)^2)$ the vacuum states contribute only to the zeroth order term $Z_v^{(0)}(\Gamma,B)$. Thus, because the operators $\hat{H}$, $\hat{P}$ and $B_0$ in the insertion into the trace of (\ref{eqn:HamiltonPartitionFunction}) all act trivially on vacuum states, the term $Z_v^{(0)}(\Gamma,B)$ can be interpreted as the Witten index counting the number of vacua with 't~Hooft flux $v$.  To be more specific, the Witten index counts the number of zero energy states taking the sign $(-1)^F$ into account, which implies that it more accurately counts the number of unpaired vacua. Unless protected by some conserved quantum number, however, any such pair of states is expected to acquire a finite energy for generic values of the parameters of the theory, implying that $Z_v^{(0)}(\Gamma,B)$ indeed counts the true number of vacua. We also note that the terms $Z_v^{(0)}(\Gamma,B)$ and $Z_v^{(1)}(\Gamma,B|\delta B)$ remain independent of $\tau$ if we expand around some connection $B$ that preserves more than one of the supersymmetries.

It follows from standard arguments that the number of states in a multiplet is invariant under a continuous change of the parameters of the theory. Furthermore, the spatial momentum ${\bf p}$ of a state obeys a quantization rule determined by the geometry of the spatial manifold and thus does not depend on the coupling constant $\tau$. For generic states, the energy $E$ in general does depend on $\tau$, but vacua and BPS-states always have $E = 0$ and $E = {\bf p}$ respectively. So $Z_v^{(0)} (\Gamma, B)$ or $Z_v^{(1)} (\Gamma, B|\delta B)$, which only receive contributions from the vacua and BPS-states respectively, are indeed independent of $\tau$. 

\subsection{Non-trivial 't~Hooft flux}
Having established this independence we can now proceed to compute $Z_v^{(0)} (\Gamma, B)$ and $Z_v^{(1)} (\Gamma, B|\delta B)$ using the results previously obtained in the weak coupling limit. For $v$ such that $\frac{1}{2}Êv \cdot v \neq 0$, $Z_v (\tau | \Gamma, B)$ vanishes identically for any $B$ in this limit, so for such 't~Hooft fluxes
\bea 
\label{eqn:ZTrivial}
Z_v^{(0)} (\Gamma, B) & = & 0 \cr
Z_v^{(1)} (\Gamma, B | \delta B) & = & 0 .
\eea
For $v \neq 0$ but $\frac{1}{2}Êv \cdot v = 0$ it is useful to rewrite $Z_v^{\rm one-loop}(\Gamma,B)$ in order to extract $Z_v^{(0)}(\Gamma,B)$ and $Z_v^{(1)}(\Gamma,B | \delta B)$, and study $S$-duality for the BPS terms. Introducing the quantity
\beq
F(S | B^4) =  \frac{| S + B^4 |}{| S |}, 
\eeq
for any $S \in H^1(M,\R)$, we can rewrite (\ref{eqn:ZvOneLoop}) (for arbitrary $B$) as
\bea
Z_v^{\rm one-loop} (\Gamma, B) & = & \prod_{\rho \in S_v} \prod_{p \in P_{\rho}} F(p | B^4) F(-p | B^4) \cr
&& \times F(p + B^1 + B^2 | B^4) F(-p + B^1 + B^2 | B^4) \cr
&& \times F(p + B^1 + B^3 | B^4) F(-p + B^1 + B^3 | B^4) \cr
&& \times F(p + B^2 + B^3 | B^4) F(-p + B^2 + B^3 | B^4) .
\eea
Expanding to linear order in $B^4$ we have (assuming that $S$ is independent of $B^4$)
\beq
F(S | B^4) = 1 + f(S | B^4) + \cO( (B^4)^2 )
\eeq
where the linear term is 
\beq
f(S | B^4) = \frac{S \cdot B^4}{| S |^2} . 
\eeq
Expanding the $\cR$-symmetry connection $B + \delta B$, imposing $B^4 = 0$, and keeping terms to linear order in $\delta B$, we thus find that
\bea 
\label{eqn:ZNonTrivial}
Z_v^{(0)} (\Gamma, B) & = & 1 \cr
Z_v^{(1)} (\Gamma, B | \delta B) & = & \sum_{\rho \in S_v} \Xi(\rho,B | \delta B) ,
\eea
where for the sake of brevity we have introduced the quantity
\bea
\Xi(\rho,B | \delta B) & = & \sum_{p \in P_{\rho}} \Bigl( f(p + B^1 + B^2 | \delta B^4) + f(-p + B^1 + B^2 | \delta B^4) \cr && + f(p + B^1 + B^3 | \delta B^4) + f(-p + B^1 + B^3 | \delta B^4) \cr && + f(p + B^2 + B^3 | \delta B^4) + f(-p + B^2 + B^3 | \delta B^4) \Bigr).
\eea
The reason for splitting the sum over $p \in P_v$ into two sums over $\rho \in S_v$ and $p \in P_{\rho}$ will become apparent below.

The BPS argument above provides a reassuring consistency check on the above result for the terms of order zero and one in $\delta B$. For this purpose it is convenient to assume that $\delta B^4 \in H^1 (M, \R / \Z)$ is directed along a cycle of $M = T^4 = \R^4 / \Gamma$ corresponding to a lattice vector $\gamma \in \Gamma$. (Any $\delta B^4$ can be arbitrarily well approximated by a multiple of a lattice vector, so by continuity this assumption is not really a restriction.) We then decompose
\beq
\Gamma = \gamma \otimes \Z \oplus \tilde{\Gamma} ,
\eeq
where $\tilde{\Gamma}$ is a rank three lattice. We interpret $\tilde{\Gamma} \otimes \R \simeq \R^3$ as space and $(\tilde{\Gamma} \otimes \R)^\perp \simeq \R$, its orthogonal complement with respect to the standard flat metric on $\R^4$,  as (Euclidean) time. The geometry of $M$ is specified by the metric on the spatial three torus $T^3 = \tilde{\Gamma} \otimes \R / \tilde{\Gamma}$ together with the spatial and temporal projections ${\bf x}$ and $t$ of the lattice vector $\gamma$ on $\tilde{\Gamma} \otimes \R$ and $(\tilde{\Gamma} \otimes \R)^\perp$ respectively. The (inverse) metric on $M$ can then be written in the form
\beq
U \cdot V = \frac{1}{t^2} (U_0 - \langle{\bf U},{\bf x}\rangle) (V_0 - \langle{\bf V},{\bf x}\rangle) + {\bf U} \cdot {\bf V} ,
\eeq
for arbitrary (Euclidean) space-time covectors $U = (U_0, {\bf U})$ and $V = (V_0, {\bf V})$ in $H^1 (M, \R)$, where $\langle \,, \rangle$ denotes the pairing between dual vector spaces
\bea
\langle \,, \rangle :  H^1(M,\R) \times H_1(M,\R) \to \R . 
\eea 
In particular, with $S = (S_0, {\bf S})$ and $\delta B^4 = (\delta B^4_0, 0)$, we have
\beq
f(S |\delta B^4) = \frac{(S_0 - \langle{\bf S},{\bf x}\rangle) \delta B^4_0}{(S_0 - \langle{\bf S},{\bf x}\rangle)^2 + t^2 {\bf S} \cdot {\bf S}} .
\eeq
A short calculation shows that this is annihilated by (the Wick rotated) Klein-Gordon operator:
\beq
(\partial_t^2 + \nabla^2_{\bf x}) f(S | \delta B^4) = 0 .
\eeq
Thus, all terms in (\ref{eqn:ZTrivial}) and (\ref{eqn:ZNonTrivial}) are annihilated by this operator. (This is of course only a non-trivial property for $Z_v^{(1)}$ and $v \neq 0, \frac{1}{2} v \cdot v = 0$.) In particular, this result is consistent with $Z_v^{(0)}(\Gamma,B)$ and $Z_v^{(1)}(\Gamma,B|\delta B)$, for arbitrary 't~Hooft flux $v \neq 0$, receiving only contributions from half-BPS states and vacua, which in Minkowski signature have $E^2 - {\bf p}^2 = 0$ and are consequently annihilated by the Klein-Gordon operator (which in general has eigenvalue $E^2 - {\bf p}^2$ for a state of energy-momentum $(E, {\bf p})$). 

\subsection{Trivial 't~Hooft flux and $S$-duality}
To summarize our results so far, we have computed the weak coupling limit of the partition functions $Z_v (\tau | \Gamma, B)$ for all $v \neq 0$ in $H^2 (M, \cC)$. Furthermore, we have argued that when $Z_v (\tau | \Gamma, B)$ (for arbitrary 't~Hooft flux $v$) is expanded around a value of $B$ which leaves one supersymmetry unbroken, the constant and linear terms, $Z_v^{(0)} (\Gamma, B)$ and $Z_v^{(1)} (\Gamma, B | \delta B)$, are in fact independent of $\tau$. When $\frac{1}{2} v \cdot v \neq 0$ and when $v \neq 0$ but $\frac{1}{2} v \cdot v = 0$, they are given by (\ref{eqn:ZTrivial}) and (\ref{eqn:ZNonTrivial}) respectively.

$S$-duality now allows us to determine the constant and linear terms $Z_0^{(0)} (\Gamma, B)$ and $Z_0^{(1)} (\Gamma, B | \delta B)$ for trivial 't~Hooft flux. We recall that the cardinality of the orbit $v \neq 0 , \frac{1}{2}v \cdot v =0$ is $n^5+n^3-n^2-1$ and note that that for every $\rho \neq 0$ there are $n^3-1$ elements in this orbit that satisfy $\rho \cdot v =0$. We also note that for fixed $\rho\neq0$ the equation $\rho \cdot v = 0$ has no solutions in the rank two orbits $\frac{1}{2}v \cdot v \neq 0$. The $S$-transformation law (\ref{eqn:STransformation}) applied to the constant and linear terms of $Z_v(\tau | \Gamma,B+\delta B)$ then yields
\bea
n^3 Z_0^{(0)} (\Gamma, B) & = & \sum_{v^\prime} Z^{(0)}_{v^\prime} (\Gamma, B) \cr
& = & Z^{(0)}_0(\Gamma,B) + \sum_{\substack{v^\prime \neq 0 \\ \frac{1}{2} v^\prime \cdot v^\prime = 0}} Z^{(0)}_{v^\prime} (\Gamma, B) + \sum_{\frac{1}{2} v^\prime \cdot v^\prime \neq 0} Z^{(0)}_{v^\prime} (\Gamma, B)  \cr
& = & Z_0^{(0)}(\Gamma, B) + (n^5 + n^3 - n^2 - 1)
\eea
and
\bea
n^3 Z_0^{(1)}(\Gamma,B|\delta B) & = & \sum_{v^\prime} Z_{v^\prime}^{(1)} (\Gamma, B|\delta B) \cr
& = & Z^{(1)}_0(\Gamma,B|\delta B) + \sum_{\substack{v^\prime \neq 0 \\ \frac{1}{2} v^\prime \cdot v^\prime = 0}} Z^{(1)}_{v^\prime} (\Gamma, B|\delta B) + \sum_{\frac{1}{2} v^\prime \cdot v^\prime \neq 0} Z^{(1)}_{v^\prime} (\Gamma,B|\delta B) \cr
& = & Z^{(1)}_0(\Gamma,B|\delta B) + \sum_{\substack{v^\prime \neq 0 \\ \frac{1}{2} v^\prime \cdot v^\prime = 0}} \sum_{\rho \in S_{v^\prime}} \Xi(\rho,B | \delta B) \cr
& = & Z^{(1)}_0(\Gamma,B|\delta B) + (n^3-1) \sum_{\rho \in S_0} \Xi(\rho,B | \delta B)
\eea
where $S_0$ is given by the obvious generalization of the set introduced in (\ref{eqn:RhoEigenvalues}). That is
\beq
\label{eqn:S0Set}
S_0 = \{ \rho \in H^1 (M, \cC) | \rho \neq 0 \} ,
\eeq
which is of cardinality $n^4 - 1$, rather than $n^2-1$ which is the cardinality of $S_v$ for $v \neq 0 , \frac{1}{2} v \cdot v = 0$. (We will consider the generalization to include also $\frac{1}{2} v \cdot v \neq 0$ below.) Thus, we have
\bea
\label{eqn:Z0NonTrivial}
Z_0^{(0)} (\Gamma, B) & = & n^2 + 1 \cr
Z_0^{(1)} (\Gamma, B | \delta B) & = & \sum_{\rho \in S_0} \Xi(\rho,B|\delta B) .
\eea

In addition to allowing the computation of $Z_0^{(0)}(\Gamma, B)$ and $Z_0^{(1)}(\Gamma, B | \delta B)$, $S$-duality also provides a consistency check of the result obtained for the constant and linear terms $Z_v^{(0)}(\Gamma,B)$ and $Z_v^{(1)}(\Gamma,B|\delta B)$ for $v \neq 0$. Using the distribution of the product $v \cdot v^{\prime}$ it is possible to verify that (\ref{eqn:ZTrivial}), (\ref{eqn:ZNonTrivial}) and (\ref{eqn:Z0NonTrivial})  are consistent with the $S$-transformation (\ref{eqn:STransformation}) also for the non-trivial 't~Hooft fluxes $v$. The details of these explicit computations are somewhat lengthy and therefor deferred to the appendix.

Thus, there is indeed a unique solution of the constraints imposed by $S$-duality for the terms $Z_v^{(0)}(\Gamma,B)$ and $Z_v^{(1)}(\Gamma,B|\delta B)$ of the partition function given by
\bea
\label{eqn:Z0Solution}
Z_v^{(0)}(\Gamma,B) & = & \left\{
\begin{array}{c c c}
n^2 + 1 & \; &v = 0 \cr
1 & \; & v \neq 0 \;,\, \frac{1}{2}v \cdot v = 0 \cr
0 & \; & \frac{1}{2} v \cdot v \neq 0 
\end{array} \right.
\eea
and
\bea
\label{eqn:Z1Solution}
Z_v^{(1)}(\Gamma,B|\delta B) & = & \left\{
\begin{array}{c c c}
{\displaystyle \sum_{\rho \in S_0} \Xi(\rho,B|\delta B)} & \; &v = 0 \cr
{\displaystyle \sum_{\rho \in S_v} \Xi(\rho,B|\delta B)} & \; & v \neq 0 \;,\, \frac{1}{2}v \cdot v = 0 \cr
0 & \; & \frac{1}{2} v \cdot v \neq 0 
\end{array} \right. \,.
\eea
A priori, the existence of this solution to the overdetermined system of constraints is not obvious, even given the independence of $\tau$ of the BPS terms and the results obtained at weak coupling for the non-trivial 't~Hooft fluxes.

However, as mentioned in the introduction, this (circumstantial) evidence for $S$-duality is hardly required for the viability of the conjecture. Rather, we attempt to illuminate the underlying structure of the duality by the following observation: As mentioned above, the set $S_v$ appearing in the weak coupling computation for the rank one 't~Hooft flux (i.e. the orbit with $v \neq 0, \frac{1}{2}v \cdot v = 0$)  has a natural generalization to arbitrary $v$. Let us define
\beq
S_v = \left\{ \rho \in H^1(M,\cC) | \rho \neq 0 , \rho \cdot v = 0 \right\}
\eeq
for any $v \in H^2(M,\cC)$. Obviously, this agrees with the previous definitions (\ref{eqn:RhoEigenvalues}) and (\ref{eqn:S0Set}) for 't~Hooft fluxes of rank zero and one. Furthermore, for $v$ in an orbit $\frac{1}{2}v \cdot v \neq 0$ there are no $\rho \neq 0$ satisfying $\rho \cdot v = 0$, so for 't~Hooft fluxes with nontrivial Pf$(v)$ we have $S_v = \emptyset$. The solutions (\ref{eqn:Z0Solution}) and (\ref{eqn:Z1Solution}) can then be expressed as
\bea
\label{eqn:ZSolution}
Z_v^{(0)} (\Gamma, B) & = & \frac{1}{n^2-1} | S_v | \cr
Z_v^{(1)} (\Gamma, B | \delta B) & = & \sum_{\rho \in S_v} \Xi(\rho,B|\delta B) 
\eea
for any value of the 't~Hooft flux $v$, where $|S_v|$ denotes the cardinality of the set $S_v$. Thus, the set $S_v$ provides an expression for the BPS terms valid for arbitrary 't~Hooft flux. Moreover, $S_v$ figures prominently in the verification of $S$-duality in the appendix, suggesting that (\ref{eqn:ZSolution}) provides a presentation of $Z_v^{(0)} (\Gamma, B)$ and $Z_v^{(1)} (\Gamma, B | \delta B)$ suitable for the purpose of studying this duality.

\section{Summary and conclusion}
To summarize, in this paper we argue that when the partition function $Z_v(\tau|\Gamma,B)$ of the $N=4$ Yang-Mills theory is expanded in a power series around an $\cR$-symmetry connection $B$ that preserve one of the four supersymmetries, the constant and linear terms $Z_v^{(0)}(\Gamma,B)$ and $Z_v^{(1)}(\Gamma,B|\delta B)$ are independent of the coupling $\tau$. For non-trivial values of the 't~Hooft flux $v$ these terms can be computed in the limit $\tau\to i\infty$ of weak gauge coupling. Furthermore, $S$-duality allows us to uniquely determine the corresponding terms also in the case of $v=0$ which is otherwise not accessible even in the weak coupling limit. The complete result for the BPS terms of the $N=4$ theory is given in (\ref{eqn:ZSolution}).

It is clear from the computations reproduced in the appendix that $S$-duality for the BPS terms is all but manifest. In particular, the $S$-transformation amounts to manipulations involving the set $S_v$, for arbitrary 't~Hooft flux $v$, and the distributions of the product $v \cdot v^\prime$. (The $T$-transformation is, as previously mentioned, already manifest in the path integral formulation.) The fact that $S$-duality acts as a reorganization of the elements in $S_v$ indicates that these sets contain all the relevant structure for the BPS terms.

A few further remarks are also in order at this point. In the expansion of the partition function, the linear part of the infinite product over $P_v$ is an infinite sum. This property appears to be crucial for the consistency with $S$-duality, since the action of the $S$-transform amounts to a reorganization of the sum over $S_v$. However, the dependences on both $\rho$ and the $\cR$-symmetry connection in the terms $\Xi(\rho,B|\delta B)$ appear to be entirely unconstrained by $S$-duality (as long as $\Xi$ is of linear order in $\delta B$ and is annihilated by the Wick rotated Klein-Gordon operator as required by the BPS-arguments) since the consistency checks involve only manipulations of $S_v$ and its defining equation $\rho \cdot v = 0$. 

The higher order terms in the power series expansion in $\delta B$, by contrast, are generally not expected to be expressible in the same way as sums over $\rho \in S_v$. Furthermore, these terms are certainly dependent on the complex coupling $\tau$. It is therefore far from obvious how to extend the results from the BPS terms to $S$-duality of the full theory.

Finally, we comment on the relationship to previous results~\cite{Henningson:2007} for the number of vacua on $\R \times T^3$ obtained in a Hamiltonian formalism: The states are then characterized by magnetic and electric 't~Hooft fluxes $m, e \in H^2 (T^3, \cC)$. (In the Hamiltonian formulation, $S$-duality interchanges $m$ and $e$, but gives no constraints on the spectrum for $m = e = 0$.) Taking $m = 0$ corresponds to a trivial gauge bundle over $T^3$. Each of the values $e \neq 0$ then corresponds to the orbit where $v \neq 0$ but $\frac{1}{2} v \cdot v = 0$, and indeed has a single vacuum. At weak coupling, these vacua can be obtained by taking linear combinations of states supported near one of the $n^3$ different flat connections over $T^3$ with holonomies given by elements of $\cC$. (Considerations of supersymmetric matrix quantum mechanics with sixteen supercharges indicate the existence of precisely one normalizable zero energy state for each such connection.) In this way, one can also construct a single vacuum with $m = e = 0$ corresponding to the orbit $v = 0$. The remaining $n^2$ vacua for this orbit are new, though. Apparently their wave-functions (at weak coupling) are not concentrated at any particular values for the holonomies of the gauge field, but rather spread out over the space of all flat connections over $T^3$. 

\vspace*{5mm}
This research was supported by grants from the G\"oran Gustafsson Foundation and the Swedish Research Council.

\newpage
\appendix

\section{Verification of $S$-duality consistency}
In this appendix we verify the consistency of the solution (\ref{eqn:ZSolution}) with the $S$-transform in some detail in order to display the underlying structure. Of course, the $S$-transform can only be considered as a consistency condition for the cases $v \neq 0$, since we are not able to compute $Z_0(\tau|\Gamma,B)$ directly even at weak coupling and must therefore rely on the $S$-transform to determine it. 

For the purpose of establishing consistency we examine the distribution of the values of the product $v \cdot v^{\prime} \in H^0(M,\R/\Z)$, recalling that $n$ is prime. First, we consider the distribution the values of $v \cdot v^{\prime}$ for the $n^5+n^3-n^2-1$ elements of the orbit $v^\prime \neq 0 , \frac{1}{2}v^{\prime} \cdot v^{\prime} = 0$ for the possible $v \neq 0$ orbits and find the result in table $\ref{tab:DistributionRankOne}$.
\begin{table}[!ht]
\beq
\begin{array}{c | c | c}
 & v \neq 0 , \frac{1}{2} v \cdot v = 0 & \frac{1}{2} v \cdot v \neq 0 \cr
\hline
v \cdot v^{\prime} = 0 &  n^4+n^3-n^2-1& n^4-1 \cr
v \cdot v^{\prime} = \frac{1}{n} & n^4 &  n^4+n^2 \cr
v \cdot v^{\prime} = \frac{2}{n} & n^4 &  n^4+n^2 \cr
\ldots & \ldots & \ldots \cr
v \cdot v^{\prime} = \frac{n - 1}{n} & n^4 & n^4+n^2 \cr
\end{array} \nonumber
\eeq
\caption{Distribution of the values of $v \cdot v^{\prime}$ for the rank one orbit $\frac{1}{2}v^{\prime} \cdot v^{\prime} = 0, v \neq 0$.}
\label{tab:DistributionRankOne}
\end{table}

Second, we consider the distribution for the $n^3-1$ 't~Hooft fluxes $v^\prime \neq 0$ satisfying $\rho \cdot v^\prime = 0$ (recall that all these are in the $\frac{1}{2}v^\prime \cdot v^\prime = 0$ orbit) for some fixed $\rho \neq 0$ and some fixed (but arbitrary) $v \neq 0$. The result is given in table \ref{tab:DistributionRhoVprimeEqualsZero}.
\begin{table}[!ht]
\beq
\begin{array}{c | c | c}
& \rho \cdot v = 0 & \rho \cdot v \neq 0\cr
\hline
v \cdot v^{\prime} = 0 & n^3- 1 &  n^2-1 \cr
v \cdot v^{\prime} = \frac{1}{n} & 0 & n^2  \cr
v \cdot v^{\prime} = \frac{2}{n} & 0 & n^2  \cr
\ldots & \ldots & \ldots \cr
v \cdot v^{\prime} = \frac{n - 1}{n} & 0 & n^2 \cr
\end{array} \nonumber
\eeq
\caption{Distribution of the values of $v \cdot v^{\prime}$ for $v^{\prime}$ satisfying $\rho \cdot v^{\prime} = 0$ for fixed $\rho \neq 0$.}
\label{tab:DistributionRhoVprimeEqualsZero}
\end{table}

Using these results we can now verify the consistency of (\ref{eqn:ZSolution}) with the transformation (\ref{eqn:STransformation}) also for the non-trivial 't~Hooft fluxes $v$. First, we consider zeroth order term of $Z_v(\tau|\Gamma,B)$ and verify 
\beq
n^3 Z_v^{(0)}(\Gamma,B) = \sum_{v^\prime} \exp \left( 2 \pi i \int_M v \cdot v^\prime \right) Z_{v^\prime}^{(0)}(\Gamma,B)
\eeq
for non-trivial 't~Hooft flux $v$.\\[2mm]
\noindent \underline{$v \neq 0 , \frac{1}{2}v \cdot v = 0$:}
\bea
{\rm LHS} & = & n^3 Z_v^{(0)}(\Gamma,B) = n^3 . \cr
{\rm RHS} & = & \sum_{v^\prime} \exp \left( 2 \pi i \int_M v \cdot v^\prime \right) Z_{v^\prime}^{(0)} (\Gamma, B) \cr
& = & Z^{(0)}_0(\Gamma,B) + \sum_{\substack{v^\prime \neq 0 \\ \frac{1}{2} v^\prime \cdot v^\prime = 0}} \exp \left( 2 \pi i \int_M v \cdot v^\prime \right)Z^{(0)}_{v^\prime} (\Gamma, B) \cr
& = & (n^2+1) + (n^4+n^3-n^2-1) + n^4 \sum_{c=1}^{n-1} \exp \left( 2 \pi i \,\frac{c}{n}\right) = n^3 .
\eea
\noindent \underline{$\frac{1}{2}v \cdot v \neq 0$:}
\bea
{\rm LHS} & = & n^3 Z_v^{(0)}(\Gamma,B) = 0 \,. \cr
{\rm RHS} & = & \sum_{v^\prime} \exp \left( 2 \pi i \int_M v \cdot v^\prime \right) Z_{v^\prime}^{(0)} (\Gamma, B) \cr
& = & Z^{(0)}_0(\Gamma,B) + \sum_{\substack{v^\prime \neq 0 \\ \frac{1}{2} v^\prime \cdot v^\prime = 0}} \exp \left( 2 \pi i \int_M v \cdot v^\prime \right)Z^{(0)}_{v^\prime} (\Gamma, B) \cr
& = & (n^2+1) + (n^4-1) + (n^4+n^2) \sum_{c=1}^{n-1} \exp \left( 2 \pi i \,\frac{c}{n}\right)  = 0 \,.
\eea
Next, we consider the linear term of $Z_v(\tau|\Gamma,B)$ and verify 
\beq
n^3 Z_v^{(1)}(\Gamma,B|\delta B) = \sum_{v^\prime} \exp \left( 2 \pi i \int_M v \cdot v^\prime \right) Z_{v^\prime}^{(1)}(\Gamma,B|\delta B) ,
\eeq
again for non-trivial 't~Hooft flux $v$.\\[2mm]
\noindent \underline{$v \neq 0 , \frac{1}{2}v \cdot v = 0$:}
\bea
{\rm LHS} & = & n^3 Z_v^{(1)}(\Gamma,B|\delta B) = n^3 \sum_{\rho \in S_v} \Xi(\rho,B|\delta B) . \cr
{\rm RHS} & = & \sum_{v^\prime} \exp \left( 2 \pi i \int_M v \cdot v^\prime \right) Z_{v^\prime}^{(1)} (\Gamma, B|\delta B) \cr
& = & Z^{(1)}_0(\Gamma,B|\delta B) + \sum_{\substack{v^\prime \neq 0 \\ \frac{1}{2} v^\prime \cdot v^\prime = 0}} \exp \left( 2 \pi i \int_M v \cdot v^\prime \right) Z^{(1)}_{v^\prime} (\Gamma, B|\delta B) \cr
& = & \sum_{\rho \in S_0} \Xi(\rho,B|\delta B) + \sum_{\substack{v^\prime \neq 0 \\ \frac{1}{2} v^\prime \cdot v^\prime = 0}} \sum_{\rho \in S_{v^\prime}} \exp \left( 2 \pi i \int_M v \cdot v^\prime \right)\Xi(\rho,B | \delta B) \cr
& = & \sum_{\rho \in S_v} \Xi(\rho,B|\delta B) + \sum_{\rho \notin S_v} \Xi(\rho,B|\delta B) + (n^3-1) \sum_{\rho \in S_v}  \Xi(\rho,B | \delta B) \cr
& & + (n^2-1) \sum_{\rho \notin S_v}  \Xi(\rho,B | \delta B) + n^2 \sum_{c=1}^{n-1} \exp\left( 2 \pi i \, \frac{c}{n}\right)\sum_{\rho \notin S_v}  \Xi(\rho,B | \delta B) \cr
& = & \left[1+(n^3-1)\right] \sum_{\rho \in S_v} \Xi(\rho,B|\delta B) + \left[1+(n^2-1)-n^2\right] \sum_{\rho \notin S_v} \Xi(\rho,B|\delta B) \cr
& = & n^3 \sum_{\rho \in S_v} \Xi(\rho,B|\delta B) .
\eea
\noindent \underline{$\frac{1}{2}v \cdot v \neq 0$:}
\bea
{\rm LHS} & = & n^3 Z_v^{(1)}(\Gamma,B|\delta B) = 0 \,. \cr
{\rm RHS} & = & \sum_{v^\prime} \exp \left( 2 \pi i \int_M v \cdot v^\prime \right) Z_{v^\prime}^{(1)} (\Gamma, B|\delta B) \cr
& = & Z^{(1)}_0(\Gamma,B|\delta B) + \sum_{\substack{v^\prime \neq 0 \\ \frac{1}{2} v^\prime \cdot v^\prime = 0}} \exp \left( 2 \pi i \int_M v \cdot v^\prime \right)Z^{(1)}_{v^\prime} (\Gamma, B|\delta B) \cr
& = & \sum_{\rho \in S_0} \Xi(\rho,B|\delta B) + \sum_{\substack{v^\prime \neq 0 \\ \frac{1}{2} v^\prime \cdot v^\prime = 0}} \sum_{\rho \in S_{v^\prime}} \exp \left( 2 \pi i \int_M v \cdot v^\prime \right)\Xi(\rho,B | \delta B) \cr
& = & \sum_{\rho \in S_0} \Xi(\rho,B|\delta B) + (n^2-1) \sum_{\rho \in S_0}  \Xi(\rho,B | \delta B) \cr
& & + n^2 \sum_{c=1}^{n-1} \exp\left( 2 \pi i \, \frac{c}{n}\right)\sum_{\rho \in S_0}  \Xi(\rho,B | \delta B) \cr
& = & \left[ 1 + (n^2-1) - n^2 \right] \sum_{\rho \in S_0} \Xi(\rho,B | \delta B) = 0 \,.
\eea


\end{document}